\documentclass[amsmath,twocolumn,superscriptaddress,prl,aps]{revtex4}
\usepackage{amsthm,amsfonts,graphicx,verbatim}
\newcommand{\be}{\begin{equation}}
\newcommand{\ee}{\end{equation}}
\newcommand{\bea}{\begin{eqnarray}}
\newcommand{\eea}{\end{eqnarray}}

\newcommand{\la}{\langle}
\newcommand{\ra}{\rangle}

\newcommand{\lp}{\left(}
\newcommand{\rp}{\right)}

\renewcommand{\phi}{\varphi}
\renewcommand{\epsilon}{\varepsilon}
\renewcommand{\vec}[1]{{\bf #1}}
\renewcommand{\Im}{{\rm Im}\,}
\renewcommand{\Re}{{\rm Re}\,}

\renewcommand{\cite}[1]{[\onlinecite{#1}]}

\begin{document}

\title{Polar Kerr Effect and Time Reversal Symmetry Breaking in Bilayer Graphene}
\author{Rahul Nandkishore}
\affiliation{Department of Physics, Massachusetts Institute of Technology, Cambridge MA 02139, USA}
\author{Leonid Levitov}
\affiliation{Department of Physics, Massachusetts Institute of Technology, Cambridge MA 02139, USA}


\begin{abstract} 
The unique sensitivity of optical response to different types of symmetry breaking can be used to detect and identify spontaneously ordered many-body states in bilayer graphene. We predict a strong response at optical frequencies, sensitive to electronic phenomena at low energies, which  arises  because of nonzero inter-band matrix elements of the electric current operator. In particular, the polar Kerr rotation and reflection anisotropy provide fingerprints of the quantum anomalous Hall state and the nematic state, characterized by spontaneously broken time reversal symmetry and lattice rotation symmetry, respectively. These optical signatures, which undergo a resonant enhancement in the near-infrared regime, lie well within reach of existing experimental techniques.
\end{abstract}

\maketitle

Optical experiments have been successfully used to probe diverse
electronic phenomena in graphene \cite{Peres}. For bilayer graphene (BLG),  physical properties such as the gate tunable bandgap \cite{yuanbo, Mak}, the band structure parameters \cite{Kuzmenko, Li, lmzhang} and the electron phonon coupling \cite{Yan,Berclaud} were investigated with the help of infrared and optical spectroscopy.
These techniques have also been used
to probe interaction effects such as band renormalization \cite{dsabergel, Tse} and exciton formation \cite{Louie, Olevano}. 
However, there has not yet been any effort to apply optical 
methods to the investigation of strongly correlated states, which are expected to form in BLG at low energies \cite{Min, Zhang10, Yang, Vafek, Nandkishore, QAH, Jung, Lemonik, longpaper}. 
This can be partly due to the low characteristic energy scales for these symmetry breaking states, estimated to be of order $1\,{\rm meV}$ \cite{Nandkishore}, which lie far outside the range of characteristic energies probed in optical experiments.

In this Letter, we point out that the problem of energy scales is offset by 
the unique sensitivity of optical response to broken symmetries. This, along with several other features, makes these methods ideally suited to the investigation of the interacting ground state of BLG. The possible broken symmetries are expected to manifest themselves through characteristic transport properties such as a non-zero Hall response or anisotropy in longitudinal conductance \cite{Min, Zhang10, Yang, Vafek, Nandkishore, QAH, Jung, Lemonik, longpaper}. Detecting these effects in transport experiments requires fabrication of samples of BLG with at least four contacts, which proves challenging in suspended BLG currently used in these experiments. 
However, optical experiments allow us to measure the AC conductivity in a contact-free manner. As we discuss below, the AC conductivity shows distinctive signatures of broken symmetry 
just like the DC conductivity. These signatures are strong due to nonzero inter-band matrix element of the current operator for 
BLG. Thus the optical response, which features additional resonant enhancement in the near-infrared regime, can be used to directly probe spontaneously broken symmetries in BLG.

A large number of possible interacting phases have been proposed for BLG \cite{Min, Zhang10, Yang, Vafek, Nandkishore, QAH, Jung, Lemonik, longpaper}. 
Recent compressibility and transport experiments on charge neutral, suspended, double gated bilayer graphene \cite{Feldman, Martin, Weitz} appear to confirm the prediction of a non-trivial interacting ground state. The experimental data was argued \cite{Weitz} to be consistent with only two of the proposed phases: the Quantum Anomalous Hall phase (QAH) predicted in \cite{QAH, Jung}, and the nematic phase predicted in \cite{Yang, Lemonik, Vafek}. Both these phases are uniquely interesting phases. The QAH phase spontaneously breaks time reversal symmetry (TRS) and 
exhibits quantum Hall effect at zero magnetic field, while the nematic state involves a distortion of the 
Dirac bandstructure that spontaneously breaks the exact rotational symmetry of the lattice. If either of these phases is confirmed in BLG, it would fulfill a long quest for an experimental realization of a QAH instability \cite{Haldane} (QAH phase) or a Pomeranchuk instability \cite{Pomeranchuk, Kivelson} (nematic phase). 

\begin{figure}
\includegraphics[width = \columnwidth]{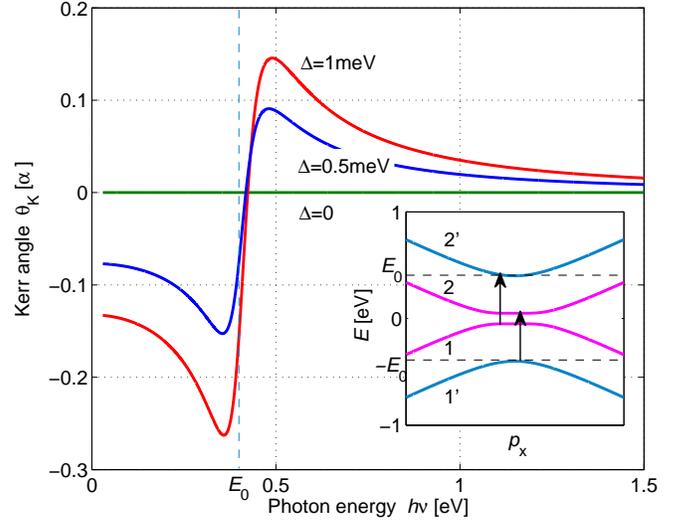} 
\caption{\label{fig: bands} Kerr angle (in units of fine structure constant $\alpha=e^2/\hbar c$) as a function of photon energy for BLG in the QAH phase. Note the resonant enhancement near $E_0=0.4\,{\rm eV}$, arising from direct transitions to the higher BLG bands, Eq.(\ref{eq:s_xy_answer}). Inset: Schematic band structure of BLG near the K point, for the QAH phase. The Kerr response arises from transitions $1'\to 2$ and  $1\to 2'$, involving states in the bands $1$ and $2$ which are affected by broken TRS. 
} 
\vspace{-6mm}
\end{figure}

Optical methods 
are ideally suited to identifying the ground state of BLG. The polar Kerr effect, wherein linearly polarized light has its polarization axis rotated upon reflection, is a well known optical probe of the Hall conductivity. It has been used to probe 
quantum Hall states  \cite{Lang}, and more recently has been applied to topological insulator thin films in the vicinity of a ferromagnet \cite{WKTse, WTse}, and to $p+ip$ superconductors \cite{Xia}. 

As we shall see, the QAH phase exhibits an AC Hall conductance in addition to the quantized DC Hall conductance, and thus the Kerr effect offers a direct test of the QAH scenario for BLG. Our analysis of optical response, taking into account transitions between four BLG bands,
reveals a resonant enhancement of the AC Hall conductivity (see Fig.\ref{fig: bands}).
This resonant enhancement occurs because the microscopic current operator has inter-band matrix elements (Fig.\ref{fig: bands} inset) corresponding to transitions from the low energy bands $1,2$ to the high energy bands $1',2'$. The resulting, resonantly enhanced Kerr rotation is plotted in Fig.\ref{fig: bands}. The predicted effect is many orders of magnitude larger than that observed in p-wave superconducting materials \cite{Xia}, and lies well within reach of existing experimental techniques. 

Optical methods can be used to probe domain formation expected to occur in the TRS breaking QAH phase. Since different domains will produce a Kerr rotation of opposite sign, the spatial domain structure can be directly imaged in optical experiments
-- a significant advantage over transport experiments, which can only measure the net effect of all domains. 
For a non-focused optical experiment, the effect of random domains will be to reduce the total Kerr angle by a factor $\sqrt{N_D}$, where ${N_D}$ is the number of domains.

While the Kerr rotation allows to test for TRS breaking, anisotropy in reflection allows to test for rotation symmetry breaking. As we discuss below, this leads to
a characteristic dependence of the reflection amplitude on the polarization angle of incident light which offers a way to test the nematic scenario for BLG \cite{Yang, Lemonik, Vafek}. 

Finally, we note that spontaneous symmetry breaking is only expected to occur below a critical temperature, estimated to be of order $1-10\,{\rm K}$ \cite{Nandkishore,Weitz}.
The optical signatures of interacting states will thus show a strong temperature dependence, and will vanish entirely above a critical temperature. This provides a way to distinguish spontaneously broken symmetries from explicitly symmetry breaking effects (e.g. magnetic impurities), which will not show any comparable temperature dependence. 

Electron properties of a clean 
BLG are governed by a four-band Hamiltonian written for the four component wavefunction $\psi=(\psi_1,\psi_2,\psi_3,\psi_4)$, describing electron wavefunction on the sublattices $A$, $B$ and $A'$, $B'$ on the two layers:
\be\label{eq:hamiltonian}
H(\vec p)=\left[\begin{array}{cccc}
         0 &  t_{\vec p} & 0 & 0\\
t_{\vec p}^* & 0 & E_0 & 0 \\
0 & E_0 & 0 & t_{\vec p} \\
0 & 0 &  t_{\vec p}^*& 0
      \end{array}
\right]
,\quad
E_0\approx 0.4\,{\rm eV}
,
\ee
with $t_{\vec p}=t_0(1+e^{-i{\vec p \vec e_1}}+e^{-i{\vec p\vec e_2}})$, where $t_0\approx 3.1 \,{\rm eV}$ is the hopping amplitude, 
and $E_0$ is bandgap parameter for the upper and lower bands. The quantity $t_{\vec p}$ vanishes at the K and K' points, behaving as $vp_+$ near point K and as $-vp_-$ near point K', where $p_\pm=p_x\pm ip_y$. 

The Hamiltonian  (\ref{eq:hamiltonian}) features four bands with energies
\be\label{eq:spectrum}
\epsilon^2(\vec p)=|t_{\vec p}|^2+\frac12E_0^2\pm \frac12E_0^2\sqrt{1+4|t_{\vec p}|^2/E_0^2}
.
\ee
Near the points K and K', this gives two massless Dirac bands $\epsilon_{1,2}(\vec p)$ that cross quadratically at zero energy, and two high-energy bands $\epsilon_{3,4}(\vec p)\approx \pm E_0$. The dispersion near K and K' can be obtained by expanding in small $t_{\vec p}/E_0$, giving $\epsilon_{1,2}=\pm |t_{\vec p}|^2/E_0 = \pm v^2 p^2/E_0$, $\epsilon_{1',2'}=\pm\lp E_0 + v^2 p^2/E_0\rp$.

We now consider the effect of interactions. Interactions can open a bulk bandgap between bands $1$ and $2$ \cite{QAH, Nandkishore, Jung, Min, Zhang10}, resulting in a bandstructure of the form Fig.\ref{fig: bands}(inset). One particularly interesting gapped state is the QAH state, \cite{QAH, Jung}, the mean field Hamiltonian of which we present below. 
To exhibit more clearly the block structure we reorder basis vectors by interchanging the components $\psi_2$ and $\psi_4$. In this representation, we obtain
\be\label{eq:hamiltonian_block}
H_{K}(\vec p, \Delta)=\left[\begin{array}{cccc}
         \Delta & 0 & v p_+ & 0\\
0 & -\Delta & 0 &  v p_-\\
v p_- & 0 & 0 & E_0 \\
0 & v p_+ & E_0 & 0
      \end{array}
\right] = H_{K'}^* (-\vec{p},-\Delta)
\ee
where $\Delta$ is the order parameter describing gap opening at the K and K' points. The other possible gapped states \cite{Min, Zhang10, Nandkishore} have a similar mean field Hamiltonian, but the sign of $\Delta$ is distributed differently among the spins and valleys. We note that under time reversal, $H_K(\Delta) \oplus H_{K'}(-\Delta) \rightarrow H_{K'}(\Delta) \oplus H_{K}(-\Delta)$, so this phase breaks TRS. In consequence, the QAH state can exhibit a non-vanishing Hall conductance at zero magnetic field. However, the gap preserves the isotropy of the bandstructure. Thus, the QAH state must exhibit isotropic longitudinal conductivity.

Next, we discuss the relation between the Hall response in the QAH phase and the Kerr rotation. We consider an experimental setup where light is incident normally on a BLG sheet that is placed on a substrate with refractive index $n$, which is taken to be real (complex case considered in \cite{supplement}). If the BLG sheet has a non-vanishing Hall conductance, then incident linearly polarized light will be reflected as elliptically polarized light, with the major axis of the ellipse rotated with respect to the incident polarization by the Kerr angle $\theta_K$. The standard formula relating the Kerr angle to the Hall conductance is $\theta_K \sim \Im(\sigma_{yx})$ \cite{White}. 
However, this formula is derived for light incident on a conducting half space, whereas we are considering a BLG sheet that is much thinner than the optical wavelength. For this case, the relationship between Hall conductivity and Kerr angle must be calculated afresh, by solving the Maxwell equations on two sides of the BLG sheet and matching solutions at the boundary. We obtain \cite{supplement}
%
\begin{equation}\label{eq: kerr angle}
\theta_K = \Re \frac{-(8\pi/c)\, \sigma_{yx}}{1 - (n+\frac{4\pi}{c}(\sigma_{xx}+i\sigma_{xy}))^2
} \approx \frac{8\pi\, \Re \sigma_{yx}}{c(n^2-1)}
. 
\end{equation}
We now calculate the magnitude of the Kerr rotation, by evaluating the conductivity. The AC conductivity can be written using the Kubo formula as
\begin{equation}\label{eq:sigma(w)}
\sigma_{xy}= \frac{e^2}{i\omega} \sum_{i,j,\vec{p}} \frac{\la i,\vec{p}| V_x | j, \vec{p} \ra \la j, \vec{p} | V_y | i,\vec{p} \ra
}{\omega - (\epsilon_{j,\vec{p}} - \epsilon_{i,\vec{p}}) + i\gamma} 
(n_{i,\vec{p}}-n_{j,\vec{p}}),
\end{equation}
%
where $i$ and $j$ are band indices, $n_{i,\vec{p}} = n(\epsilon_{i,\vec{p}})$ is a Fermi function, and the sum over momenta $\vec{p}$ stands for an integral. The velocity operators $V_{\alpha}$ are defined as $V_{\alpha}=\partial H(\vec p)/\partial p_{\alpha}$, and $\gamma$ describes the excited state lifetime.

We focus on the contributions which correspond to optical interband transitions between the massless low energy bands (i=1,2), and the high energy bands (i=1',2'), which are separated from the low energy bands by the energy $E_0$. We focus on these transitions because they are of resonant character at a frequency close to the band separation energy $E_0$, and hence dominate the optical response. We now note that $\la i,\vec{p}| V_{\alpha} | j, \vec{p} \ra \la j, \vec{p} | V_{\beta} | i,\vec{p} \ra = {\rm Tr} \big(V_{\alpha} \Pi _{i,\vec{p}} V_{\beta} \Pi_{j,\vec{p}} \big)$, where $\Pi_{i,\vec{p}}$ projects onto the state in band $i$ with momentum $\vec{p}$. 
%
%
%
Assuming we are at a temperature $T\ll E_0/k_B = 4000K$, we obtain
\bea\nonumber 
\sigma_{xy}(\omega)=&&\frac{e^2}{i\omega} \int \frac{d^2p}{(2\pi)^2}\frac{{\rm Tr}\lp V_{x} \Pi_{1'} V_{y} \Pi_{2} +V_{x} \Pi_{1} V_{y} \Pi_{2'}\rp}{\omega+\epsilon_{1'}(\vec p)-\epsilon_2(\vec p)+i \gamma} 
\\\label{eq:sigma(w)_Tr}
&& - \frac{{\rm Tr}\lp V_{x} \Pi_{2} V_{y} \Pi_{1'} +V_{x} \Pi_{2'} V_{y} \Pi_{1}\rp}{\omega-\epsilon_{1'}(\vec p)+\epsilon_2(\vec p)+i \gamma} \label{eq: traceH}
,
\eea
where we used the relation $\epsilon_{2'}-\epsilon_1=\epsilon_2-\epsilon_{1'}$ 
that follows from particle/hole symmetry of the Hamiltonian (\ref{eq:hamiltonian_block}).

We evaluate the expression (\ref{eq:sigma(w)_Tr}) for $\vec p$ near point K with the help of the projectors
%
\be
\Pi_{1,2}=\frac12\lp 1\pm \frac{h(\vec p)}{||h(\vec p)||}\rp,\quad
\Pi_{1',2'}=\frac{1\pm\tilde\sigma_x}2
\ee
Here $\Pi_{1',2'}$ project on the $B1$ and $A2$ sublattices (lower right corner of the Hamiltonian in (\ref{eq:hamiltonian_block})), and $\tilde \sigma_x$ acts on this subspace. Meanwhile, $\Pi_{1,2}$ project on the $A1$ and $B2$ sublattices (upper left corner of Eq.(\ref{eq:hamiltonian_block})), and $h(\vec{p})$ is the effective two band Hamiltonian for the massless Dirac states, which 
 has eigenvalues $E(\vec{p}) = \pm ||h(\vec{p})||$. 
The trace over projectors takes the form 
\bea\nonumber 
g^{1'2}_{\alpha\beta}&=& {\rm Tr}\lp V_\alpha \Pi_{1'} V_\beta \Pi_2\rp=\la 1' | V_{\beta} \Pi_2 V_{\alpha}|1' \ra
\\
&=& \frac14 \left[\begin{array}{c}
\nabla_\beta t^*(\vec p)\\ 
\nabla_\beta t(\vec p)
\end{array}
\right]^{\rm T}
\lp 1-\frac{h(\vec p)}{||h(\vec p)||}\rp \left[\begin{array}{c}
\nabla_\alpha t(\vec p)\\
\nabla_\alpha t^*(\vec p)
\end{array}
\right]   \label{eq: traces}\\\nonumber
g_{\alpha \beta}^{12'} &=& {\rm Tr}\lp V_\alpha \Pi_1 V_\beta \Pi_{2'}\rp=\la 2' |V_{\alpha} \Pi_1V_{\beta}|2' \ra
\\\nonumber 
&& = \frac14 \left[\begin{array}{c}
\nabla_\alpha t^*(\vec p)\\
-\nabla_\alpha t(\vec p)
\end{array}
\right]^{\rm T}
\lp 1+\frac{h(\vec p)}{||h(\vec p)||}\rp \left[\begin{array}{c}
\nabla_\beta t(\vec p)\\
-\nabla_\beta t^*(\vec p)
\end{array}
\right].
\end{eqnarray}
Here $\nabla_{\alpha}$ denotes $\partial/\partial p_{\alpha}$.
We now compute $h(\vec{p})$ 
by using second order perturbation theory in $vp/E_0$, and obtain 
\be\label{eq:h(p)}
h_K(\vec p)=\left[\begin{array}{cc}
\Delta & v^2 p_+^2/E_0\\
 v^2 p_-^2/E_0 & -\Delta
\end{array}
\right], \quad h_{K'}(\Delta) = h_K^*(-\Delta).
\ee
This result agrees with \cite{McCann}. We substitute this two band Hamiltonian into Eq.(\ref{eq: traces}) 
and obtain 
%
\begin{eqnarray}
g^{1'2}_{xy}= g^{12'}_{xy} &=&\frac14\lp 1-\frac{\Delta}{||h(\vec p)||}\rp\nabla_y t^*(\vec p)\nabla_x t(\vec p)
\\\nonumber
&& + \frac14\lp 1+\frac{\Delta}{||h(\vec p)||}\rp\nabla_y t(\vec p)\nabla_x t^*(\vec p) 
,
\eea
where we suppressed the terms arising from off-diagonal parts of $h(\vec p)$ --- these terms give zero upon integration over $d^2p$. 
Hence, 
we find 
$g^{12'}_{xy}=-g^{2'1}_{xy}=\frac12 iv^2\frac{\Delta}{||h(\vec p)||}$. We substitute these results into Eqs.(\ref{eq: traceH}),(\ref{eq: traces}), to obtain
%
%
%
\be
\sigma_{xy}(\omega)=\frac{N e^2v^2 \Delta}{(2\pi)^2 \omega}\int \frac{d^2p}{||h(\vec p)||} \left[\frac{1}{\omega+ i \gamma-\Omega_{\vec{p}}}  + (\Omega_{\vec{p}} \rightarrow - \Omega_{\vec{p}})\right]
\ee
where $N=4$ is the number of spin/valley flavors, and $\Omega_{\vec{p}} = \epsilon_2(\vec{p}) - \epsilon_{1'}(\vec{p})$. 

We now specialize to optical frequencies $\omega \gg \Delta$, and also assume $\gamma \gg \Delta$. We approximate by taking $||h(\vec{p})|| \approx v^2 p^2 /E_0$ and $\Omega_p \approx E_0 + 2 v^2 p^2/ E_0$, and perform the momentum integral in polar co-ordinates. The log divergence near $p^2 = 0$ is cut by $|\Delta|$, but there is no need for any high energy cutoff. In this manner, we obtain the Hall conductivity 
%
\begin{eqnarray}\label{eq:s_xy_answer}
\sigma_{xy}(\omega) &=& \frac{N e^2}{h} \frac{\Delta}{2\omega}\bigg[\frac{E_0}{\omega + E_0 + i\gamma}\ln \bigg(\frac{E_0 +\omega + i\gamma}{2|\Delta|}\bigg) \nonumber\\&+&\frac{E_0}{\omega - E_0 + i\gamma}\ln \bigg(\frac{E_0 -\omega - i\gamma}{2|\Delta|}\bigg) \bigg]
.
\end{eqnarray}
%
%
There is also a contribution from $1\rightarrow2$ transitions, which may be evaluated in the two band model \cite{longpaper, sinner}. This contribution, 
extrapolated to optical frequencies $\omega\sim E_0\gg\Delta$ is of order $(\Delta |\Delta|/\omega^2)e^2/h$. This is smaller than the contribution (\ref{eq:s_xy_answer}) by a large factor 
\begin{equation}
\frac{E_0}{\Delta}\ln\frac{E_0}{\Delta}\gg 1
. \label{eq: factor}
\end{equation}
Thus, the Hall conductivity at optical frequencies is dominated by transitions to the higher bands, necessitating our four band analysis.

From the result Eq.(\ref{eq:s_xy_answer}) and the expression Eq.(\ref{eq: kerr angle}) we can extract the Kerr angle $\theta_K$. We take $n=1.5$, which describes ${\rm SiO_2}$ substrate, and take $\Delta \approx 10^{-3}\,{\rm eV}$ \cite{Nandkishore}, 
and $\gamma = 0.05\,{\rm eV}$. The resulting Kerr angle as a function of frequency is plotted in Fig.\ref{fig: bands}. In optical experiments on cuprate materials, Kerr angles as small at $10^{-9}$ have been measured \cite{Xia}. The six orders of magnitude larger Kerr rotation in the QAH phase should thus 
be comfortably within reach of experiments. 

The nematic state \cite{Yang, Lemonik, Vafek} is another interesting ordered state proposed to explain the experiments \cite{Weitz, Martin}. This state is time-reversal invariant, featuring no Kerr rotation. Instead, it breaks rotation symmetry of graphene crystal lattice. The Hamiltonian for this state is 
%
\be\label{eq:hamiltonian_block_nem}
H_K(\vec p, \Delta)=\left[\begin{array}{cccc}
         0 & \Delta & v p_+ & 0\\
\Delta^* & 0 & 0 &  v p_-\\
v p_- & 0 & 0 & E_0 \\
0 & v p_+ & E_0 & 0
      \end{array}
\right] = H_{K}^*( -\vec{p},\Delta)
. 
\ee
After reduction to the two low-energy bands, it becomes
\begin{equation}
h_K(\vec p, \Delta)=\left[\begin{array}{cc} 0 & \frac{v^2 p_+^2}{E_0} + \Delta\\
\frac{ v^2 p_-^2}{E_0} + \Delta^*& 0
\end{array}
\right] = h_K^*(-\vec{p}, \Delta).
\end{equation}
This Hamiltonian describes splitting of the quadratic band crossing into two linear band crossings. The argument of the nematic order parameter $\Delta$ specifies the orientation of the nematic axis, which is defined as the line joining the two linear band crossings. The nematic axis makes an angle $\phi = -\pi/2 + \arg(\Delta)/2$ with respect to the $p_x$ axis. 
The nematic state manifestly breaks the approximate rotation invariance of the low energy band-structure, which manifests itself in an anisotropic longitudinal conductivity. Writing $\sigma(\theta) = \sigma_0 + \delta \sigma(\theta)$, where $\theta$ is the angle with respect to the $x$ axis, we obtain an expression for the reflection amplitude $r(\theta)$, 
%
\begin{equation}
r(\theta) \approx \frac{1-n}{n+1} - \frac{8\pi }{c(n+1)^2} \delta \sigma(\theta). 
\end{equation}
For high frequencies $\omega \gg \Delta$, we calculate using the formalism introduced above that 
\begin{equation}
\delta \sigma (\theta) \sim \frac{i e^2}{\hbar} \frac{|\Delta|}{\omega} \ln \frac{E_0 - \omega - i\gamma}{\Delta} \cos(2(\theta - \phi))  \label{eq: delta sigma}
\end{equation}
Again, this exceeds the anisotropy calculated in the two band model \cite{longpaper} by the large factor Eq.(\ref{eq: factor}). 
We note that trigonal warping of the BLG bandstructure arising from higher neighbor hopping can also lead to a reflection anisotropy. However, these effects respect the threefold rotation symmetry of the lattice. In contrast, the anisotropy resulting from formation of a nematic state exhibits a twofold rotation symmetry. The  breaking of the exact lattice rotation symmetry 
can serve as diagnostic of the nematic state.

To conclude, optical experiments can be used to probe broken symmetries in BLG by measuring the conductivity in a contact free manner. The polar Kerr effect, by providing a means for measuring Hall conductivity, can be used to detect the QAH phase. TRS breaking gapped states that do not display a Hall conductance \cite{Jung} can also be probed using the Kerr effect, although for these states the Kerr angle will be smaller than that for the QAH state by the small parameter $d/\lambda$, where $d = 3 \AA$ is the BLG interlayer spacing and $\lambda$ is the wavelength of the light used in the experiment \cite{Dzyaloshinskii}. Nevertheless, this much weaker Kerr rotation will still be much larger than that measured in \cite{Xia}, and will be within reach of experiments. Meanwhile, the nematic scenario for BLG may be probed by looking for an angle dependence of the reflection amplitude, which provides a direct test of broken rotational symmetry. 

We acknowledge useful conversations with Jing Xia. This work was supported by Office of Naval Research Grant No. N00014-09-1-0724.

\section{Appendix}
We consider light incident normally on a BLG sheet placed on a substrate with refractive index $n = n' + i n''$. Incident and transmitted waves propagate in the $+z$ direction, while the reflected wave propagates in the $-z$ direction. The BLG sheet is taken to be in the $z=0$ plane, whereas the substrate occupies the halfspace $z>0$. We calculate the reflection  amplitudes for incident light that is linearly polarised along the $x$ axis. 
The reflected wave, $E_{r} = r_{xx} \hat{\vec x} + r_{yx} \hat{\vec y}$, is linearly polarised if $\Im(r_{yx}/r_{xx}) = 0$, and elliptically polarised otherwise. The major axis of the polarisation is rotated with respect to the x axis by the Kerr angle $\theta_K = \Re(r_{yx}/r_{xx})$. 

We start with rewriting Maxwell's equations $\vec k\times\vec H=-\frac{\omega}{c}\vec D$, $\vec k\times\vec E=\frac{\omega}{c}\vec H$ as 
\begin{eqnarray}
k H_x &=& - \frac{\omega}{c} n^2 E_y \qquad k H_y =  \frac{\omega}{c} n^2 E_x \\ 
k E_x &=&  \frac{\omega}{c}  H_y \qquad k E_y =  - \frac{\omega}{c} H_x \nonumber
\end{eqnarray}
%
 Similar relations hold in the vacuum region with $n$ replaced by $1$. At the interface $z=0$ we must match EM field amplitudes on both sides using continuity of the $E$ field $E_< = E_> $ and the Amp\`ere's law for the $H$ field:
\[
\big(H_> - H_<\big)_x = \frac{4\pi}{c} \big(\sigma E)_y \qquad \big(H_> - H_<\big)_y =  - \frac{4\pi}{c} \big(\sigma E)_x.
\]
%

For an incident wave polarised along the $x$ axis, $E_{\rm in} = \hat{\vec x}$
, we have $E_>= t_{xx} \hat{\vec x} + t_{yx} \hat{\vec y}$, $E_<= (1+r_{xx}) \hat{\vec x} + r_{yx} \hat{\vec y}$, $H_>=  n (t_{xx} \hat{\vec y} - t_{yx} \hat{\vec x})$, $H_<= (1 - r_{xx}) \hat{\vec y} +  r_{yx} \hat{\vec x}$.
Applying Amp\`ere's law to the magnetic field leads to the continuity relations 
\begin{eqnarray}
n t_{xx} - 1 + r_{xx} = - \frac{4\pi}{c} \big(\sigma_{xx} t_{xx} + \sigma_{xy} t_{yx}\big) \nonumber \\
nt_{yx} + r_{yx} = -  \frac{4\pi}{c} \big(\sigma_{yx} t_{xx} + \sigma_{yy} t_{yx}\big) 
\end{eqnarray}
Eliminating $t$ using the continuity relations for electric field, $E_<=E_>$, 
we obtain the single matrix equation
\begin{equation}
\left[ \begin{array}{cc} n+1+\frac{4\pi }{c} \sigma_{xx} & \frac{4\pi }{c} \sigma_{xy} \\ \frac{4\pi}{c} \sigma_{yx} & n + 1 + \frac{4\pi}{c} \sigma_{yy} \end{array} \right] \left[ \begin{array}{c} r_{xx} \\ r_{yx} \end{array} \right] = \left[\begin{array}{c} 1 - n - \frac{4\pi }{c} \sigma_{xx} \\  -\frac{4\pi }{c} \sigma_{yx} \end{array}\right]
\end{equation}
This equation can be solved to obtain
\begin{equation}
\left[ \begin{array}{c} r_{xx} \\ r_{yx} \end{array} \right] = \frac{1}{D} \left[ \begin{array}{c} 1 - (n + \frac{4\pi}{c} \sigma_{xx})^2 + (\frac{4\pi}{c})^2 \sigma_{xy} \sigma_{yx} \\   - \frac{8\pi}{c} \sigma_{yx} \end{array}\right]; \label{eq: kerr rotation}
\end{equation} 
We have denoted $D = (n+1+\frac{4\pi}{c} \sigma_{xx})^2
 - |\frac{4\pi}{c}\sigma_{xy}|^2 \approx (n+1)^2$, and have assumed isotropy, so that $\sigma_{xx} = \sigma_{yy}$. The Kerr angle is given by $\Re(r_{yx}/r_{xx})$, and thus takes the form 

\begin{eqnarray}
\theta_K &=& \Re \bigg[\frac{- \frac{8\pi}{c} \sigma_{yx}}{1 - (n+\frac{4\pi}{c}\sigma_{xx})^2 + (\frac{4\pi}{c} |\sigma_{xy}|)^2}\bigg]\nonumber\\ &\approx& \Re \bigg( \frac{8\pi}{c(n^2-1)} (\sigma_{yx}) \bigg)\nonumber\\
&=& \frac{8\pi [(n'^2 - n''^2 - 1) \Re(\sigma_{xy}) - 2 n'n'' \Im(\sigma_{xy})]}{c[(n'^2 - n''^2 - 1)^2 + 4 n'^2 n''^2]}\nonumber
\end{eqnarray}
where in the last line we have taken $n = n' + i n''$ with $n'$ and $n''$ real. Now if we assume $n = n'$ and $n'' = 0$, we obtain the formula quoted in the main text
\begin{equation}
\theta_K = \frac{8\pi}{c(n^2-1)} \Re \sigma_{xy}
\end{equation}

\begin{thebibliography}{99}
\vspace{-7mm}

\providecommand{\natexlab}[1]{#1}
\providecommand{\url}[1]{\texttt{#1}}
\expandafter\ifx\csname urlstyle\endcsname\relax
  \providecommand{\doi}[1]{doi: #1}\else
  \providecommand{\doi}{doi: \begingroup \urlstyle{rm}\Url}\fi

\bibitem[Peres (2010)]{Peres}
N. M. R. Peres, Rev. Mod. Phys. {\bf 82}, 2673 (2010).

\bibitem[yuanbo (2009)]{yuanbo}
Y. Zhang \emph{et al} 
Nature, {\bf 459}, 820 (2009).

\bibitem[Mak (2009)]{Mak}
K. F. Mak, C. H. Lui, J. Shan and T. F. Heinz, Phys. Rev. Lett. {\bf 102}, 256405 (2009).

\bibitem[Kuzmenko (2009)]{Kuzmenko}
A. B. Kuzmenko \emph{et al},
Phys. Rev. B {\bf 79}, 115441 (2009).

\bibitem[Li (2009)]{Li}
Z. Q. Li \emph{et al},
Phys. Rev. Lett. {\bf 102}, 037403 (2009).

\bibitem[lmzhang (2008)]{lmzhang}
L. M. Zhang \emph{et al}, 
Phys. Rev. B {\bf 78}, 235408 (2008).

\bibitem[Yan (2008)]{Yan}
J. Yan, E. A. Henriksen, P. Kim and A. Pinczuk, Phys. Rev. Lett. {\bf 101}, 136804 (2008).

\bibitem[Berclaud (2010)]{Berclaud}
S. Berclaud, M. Y. Han, K. F. Mak, L. E. Brus, P. Kim and T. F. Heinz, Phys. Rev. Lett. {\bf 104}, 227401 (2010).

\bibitem[dsabergel (2011)]{dsabergel}
D. S. L. Abergel and T. Chakraborty, Nanotechnology {\bf 22}, 015203 (2011).

\bibitem[Tse (2009)]{Tse}
W. Tse and A. H. MacDonald, Phys. Rev. B {\bf 80}, 195418 (2009).

\bibitem[Louie (2009)]{Louie}
L. Yang, J. Deslippe, C. H. Park, M. L. Cohen and S. G. Louie, Phys. Rev. Lett. {\bf 103}, 186802 (2009).

\bibitem[Olevano (2010)]{Olevano}
P. E. Trevisanutto, M. Holzmann, M. Cote and V. Olevano, Phys. Rev. B. {\bf 81}, 121405(R) (2010).

\bibitem[Min (2008)]{Min}
H. Min, G. Borghi, M. Polini and A.H. MacDonald, Phys. Rev. B 77, 041407(R) (2008).

\bibitem[R. Nandkishore and L. Levitov (2009)]{Nandkishore}
R. Nandkishore and L. Levitov. Phys. Rev. Lett. {\bf 104}, 156803 (2010).

\bibitem[Zhang (2010)]{Zhang10} F. Zhang, H. Min, M. Polini, and A. H. MacDonald
Phys. Rev. B {\bf 81}, 041402(R) (2010).

\bibitem[QAH (2010)]{QAH}
R. Nandkishore and L. Levitov, Phys. Rev. B {\bf 82}, 115124 (2010).

\bibitem[Jung (2010)]{Jung}
J. Jung, F. Zhang and A. H. MacDonald, Phys. Rev. B 83, 115408 (2011).

\bibitem[Vek (2009)]{Yang}
O. Vafek and K. Yang, Phys. Rev. B {\bf 81}, 041401(R) (2010).

\bibitem[Lemonik (2010)]{Lemonik}
Y. Lemonik, I. L. Aleiner, C. Toke and V. I. Fal'ko, Phys. Rev. B {\bf 82}, 201408(R) (2010).

\bibitem[Vafek (2010)]{Vafek}
O. Vafek, Phys. Rev. B {\bf 82}, 205106 (2010).

\bibitem[phenomenology (2011)]{longpaper}
R. Nandkishore and L. Levitov, arXiv: 1002.1966v2 (to appear)

\bibitem[Feldman (2009)]{Feldman}
B. Feldman, J. Martin and A. Yacoby, Nature Physics {\bf 5}, 889 (2009).

\bibitem[Feldman (2010)]{Martin}
J. Martin, B. E. Feldman, R. T. Weitz, M. T. Allen and A. Yacoby, Phys. Rev. Lett. {\bf 105}, 256806 (2010).

\bibitem[Weitz (2010)]{Weitz}
R. T. Weitz, M. T. Allen, B. E. Feldman, J. Martin and A. Yacoby, Science {\bf 330}, 812 (2010).

\bibitem[Haldane (1988)]{Haldane}
F. D. M. Haldane, Phys. Rev. Lett. {\bf 61}, 2015 (1988).

\bibitem[Pomeranchuk (1958)]{Pomeranchuk}
I. J. Pomeranchuk, Sov. Phys. JETP {\bf 8}, 361, (1958).
 
\bibitem[Kivelson (2010)]{Kivelson}
E. Fradkin, S. A. Kivelson, M. J. Lawler, J. P. Eisenstein and A. P. Mackenzie, Annu. Rev. Condens. Matter Phys., {\bf 1}, 153 (2010).

\bibitem[Lang (2005)]{Lang}
R. Lang {\it et al}, Phys. Rev. B {\bf 72}, 024430 (2005).

\bibitem[WKTse (2010)]{WKTse}
W.Tse and A. H. MacDonald, Phys. Rev. Lett. {\bf 105}, 057401 (2010).

\bibitem[WTse (2010)]{WTse}
W. Tse and A. H. MacDonald, Phys. Rev. B {\bf 82}, 161104(R) (2010)

\bibitem[Xia (2006)]{Xia}
J. Xia, Y. Maeno, P. T. Beyersdorf, M. M. Fejer and A. Kapitulnik, Phys. Rev. Lett. {\bf 97}, 167002 (2006).

\bibitem[White (1979)]{White}
R. M. White and T. H. Geballe, \emph{Long Range Order in Solids}, pp. 317, 321 [Academic Press (1979)].

\bibitem[supplement (2011)]{supplement}
See Appendix 


\bibitem[Mc Cann and Fal'ko(2006)]{McCann}
E. McCann and V. Fal'ko, Phys. Rev. Lett. {\bf 96}, 086805 (2006).

\bibitem[Ziegler (2010)]{sinner}
A. Hill, A. Sinner and K. Ziegler, cond-mat: 1005.3211v1 (2010), New. J. Phys. 13, 035023 (2011)

\bibitem[Dzyaloshinskii (1995)]{Dzyaloshinskii}
I. Dzyaloshinskii and V. Papamichail, Phys. Rev. Lett. {\bf 75}, 3004 (1995).
\end{thebibliography}
\end{document}